\documentclass{appolb}
\usepackage{epsfig}
\begin{document}
\pagestyle{plain}
\newcount\eLiNe\eLiNe=\inputlineno\advance\eLiNe by -1

\title{On the anomalous CP violation and 
noncontractibility of the physical space}
\author{Davor Palle}
\address{Zavod za teorijsku fiziku, Institut Rugjer Bo\v skovi\' c \\
Bijeni\v cka cesta 54, 10000 Zagreb, Croatia}
\maketitle
\begin{abstract}
There is a growing evidence for the anomalously large
semileptonic CP asymmetry in the B meson system measured at the Tevatron.
The noncontractible space, as an alternative symmetry-breaking
mechanism to the Higgs mechanism, can change standard field theoretic calculations of the physical processes mediated through quantum loops
for large external momenta or large internal masses.
The presence of the W bosons and t-quarks in loops of the B meson
mixing can enhance the corresponding semileptonic CP asymmetry when
the loop integration is up to the universal Lorentz and gauge invariant
UV cut-off.
We show that the enhancement is roughly 13\%, thus the possible
deviation is measurable at the Tevatron, LHCb, SuperKEKB and SuperB facilities. 
\end{abstract}

\section{Introduction and motivation}

Any theory beyond the Standard Model (SM) in particle physics
must fulfil very severe theoretical and experimental 
requirements.
Let us enumerate the most important ones: 1) presence of massive
neutrinos and a very probable CP violation in the lepton sector,
2) it must contain a candidate for a cold dark matter particle, 
 3) the theory should explain
the existence of only three fermion families,
mass hierarchy and very small neutrino masses,
4) the SM CP violation in the baryon sector is insufficient to
generate a very large excess of baryons over antibaryons in the
Universe, 5) the theory of elementary particles should have
some fundamental relationship to the theory of gravity.

One can also find the theoretical attempts where the dark matter
problem is resolved with the modified theories of gravity
without any dark matter particle.
However, the rotational curves of galaxies, gravity lensing by dark halos,
fluctuations of the CMBR, large scale galaxy and cluster correlations,
Sunyaev-Zeldovich effect, age of the Universe, etc. are phenomena
that are very difficult
to simultaneously explain without the introduction of dark matter
\cite{Kolb}.

Recent Tevatron results \cite{D0} for the semileptonic CP
asymmetry of $B_{d}$ and $B_{s}$ mesons reveal much larger
baryon CP violation than the SM expectations. They induced
a lot of theoretical attempts with beyond the SM physics in
order to reproduce the experimental data. 

In this letter we
want to investigate the impact of the new symmetry-breaking
mechanism on the most sensitive electroweak observables.
The theory proposed in \cite{Palle1} is formulated 
to understand and contrive the relations between gauge, conformal
and discrete symmetries in particle physics and 
solving the problem of the unitarity ($SU(2)$ global anomaly) 
and the problem of the UV (zero distance) infinity.

The acronym BY denotes precisely the attributes of the theory with
respect to two previously mentioned issues:
"A" theory contains Nambu-Goldstone scalars without carrying the lepton
number, while "B" theory contains Nambu-Goldstone scalars with the lepton
number.
"X" denotes a theory with the Higgs mechanism,"Y" denotes a theory of
the noncontractible space, i.e., the existence of the universal UV-cutoff.
Thus, the four possible combinations are AX, BX, AY and BY. The AX theory
is the SM, while our favourite theory BY is trying to elucidate
two essentially nonperturbative problems: the $SU(2)$ global anomaly and
the UV singularity. 

The universal gauge and Lorentz invariant UV-cutoff
in the spacelike domain of the Minkowski spacetime
$\Lambda=\frac{\hbar}{c d}=\frac{2}{g}\frac{\pi}{\sqrt{6}}
M_{W} \simeq 326 GeV$, g=weak coupling, $M_{W}$=weak boson mass,
is fixed by the trace anomaly \cite{Palle1}.
It is the only parameter of this symmetry-breaking mechanism.

Let us briefly summarize the particle content and the most
important phenomenological consequences of the BY theory:

-Light Majorana neutrinos as a hot dark matter component and
heavy Majorana neutrinos as a cold dark matter component
of the Universe are $SU(3)$ singlet fermions of the $SU(3)$
conformal unification scheme of the BY theory. The absence
of the Higgs scalar appears crucial for the cosmological
stability of the heavy Majorana neutrinos as cold
dark matter particles \cite{Palle2} which decay to the pairs
of the weak interaction gauge bosons \cite{Palle2a}.

-A study of Dyson-Schwinger equations of the Abelian model
in the first nontrivial approximation within UV-finite
theory gives mass functions of fermions with the magnitudes close
to that observed in nature. Since any mass function has
the form $m(q^{2},\alpha_{g},\Lambda)=\Lambda f(q^{2},\alpha_{g})$,
where $q^{2}$ is Lorentz invariant squared momentum,
$\alpha_{g}=g^{2}/4\pi$ is the coupling and $\Lambda=326 GeV$
is the UV-cut-off defined by the weak boson masses and the weak coupling,
if the universal UV cut-off exists, it must be fixed by weak
interactions \cite{Palle3}. The appearance of multinode solutions
is a general phenomenon of bootstrap equations that helps to
resolve the number and the mass gaps between the fermion families
\cite{Palle3}.

-The lepton CP violating phase and the dynamic of the 
heavy Majorana neutrinos, which are strongly coupled to the
Nambu-Goldstone scalars, suffice to provide enough power to
generate lepton asymmetries in the early Universe. Light Majorana
neutrinos, on the other hand, induce vorticity of the Universe
with right-handed chirality \cite{Palle4} within the Einstein-Cartan
cosmology. The resulting angular momentum of the Universe can play the 
role of the dark energy within the nonsingular
Einstein-Cartan cosmology \cite{Palle5}.

Since the BY theory has very appealing features such as the existence of the mixed
light neutrinos and the dark matter particle, we enumerate some additional
phenomenological problems based on recent experiments that could be explained 
within the UV-finite BY theory.

The Tevatron experiments reported already \cite{TeV1}
larger cross sections than predicted by the standard field
theoretic QCD (see also Tevatron papers quoted in \cite{Palle6}).
Although, the shape of the cross sections at large scales ($\mu >
250 GeV$) could be reproduced by the SM QCD, the magnitude of
the cross section quotient $\sigma(630 GeV)/\sigma(1800 GeV)$
is more than $10\%$ away from the SM prediction.
This is in accord with the prediction of the
BY theory that the QCD in the noncontractible space is not
an asymptotically free gauge theory $\lim_{\mu \rightarrow \infty}
\alpha_{s}^{\Lambda}(\mu) \neq 0$ \cite{Palle6}.
To one loop order, the quotient between the BY ($\Lambda$) and the SM 
($\infty$) strong couplings can be evaluated \cite{Palle6}:
$\alpha_{s}^{\Lambda}/\alpha_{s}^{\infty}(\mu=1; 2; 3.5; 7\ TeV)=
1.23; 1.31; 1.38; 1.47$, respectively.

The observed forward-backward t-quark pair asymmetry at the Tevatron
\cite{FBasymm} deviates substantially from the theoretical prediction
\cite{Kuehn}. A larger QCD coupling of the BY theory and
the deviation of the corresponding box diagram from the standard
field theoretic QCD estimate, could improve the agreement with
the experimental value because the asymmetry is proportional
to $\alpha_{s}$. The systematic errors are reduced because the
charge asymmetry is defined as a quotient of the difference and
the sum of the forward and backward integrated cross sections.
The observed enhancement \cite{FBasymm} is nonresonant with
larger deviations from the SM QCD for larger invariant masses of
the t-quark pairs. The huge enhancement of the asymmetry within
the BY theory is confirmed in \cite{Palle7}.

The branching ratio for the rare decay $B_{s}\rightarrow \mu\mu$ appears
 to be lower for more than $30\%$ in the BY theory compared
with the SM \cite{Palle8}. The LHCb could measure this mode very soon
\cite{LHCbBmu}.

High energy hadron colliders, such as the LHC and the Tevatron,
require very demanding analyses of data with incorporated 
SM physics. Rediscovering the SM physics, the LHC is faced 
with the phenomena that can be attributed to the leading order as
problems with the power of the QCD coupling at
certain scales such as: energy flow \cite{flow},
$B^{+}$ production \cite{Bplus}, bottomonium production
\cite{bott}, $J/\Psi$ production \cite{JPsi},
high energy dijets \cite{dijets}, multiplicities of charged
particles \cite{multipli}, diphoton production \cite{diphoton}, etc. These problems call for the
reevaluation of the "background" physics with the BY theory.

In the next section 
we are concentrated on the evaluation of the loop dominated
mixing of the neutral mesons within the BY theory:
$SU(3)\times SU(2)_{L}\times U(1)$ gauge theory within noncontractible 
space ($\Lambda < \infty$) without the Higgs scalar.

\section{CP violation and the B meson mixing} 

The BY theory differs from the SM in the lepton sector 
already at
the tree level because of the heavy Majorana neutrinos
\cite{Palle1},
but in the quark sector, one has to search for electroweak
processes mediated by quantum loops such as rare decays
or CP violated transitions. It would be instructive
to compare possible modifications 
 to the CP
violated mixings of the K and B mesons, which are
due to the universal Lorentz and gauge invariant  
UV cut-off $\Lambda\simeq 326 GeV$.

The formalism of meson mixing is very well known from
the old studies of the strange quark physics, but here we 
refer to the updated analysis of $B_{s}-\bar{B_{s}}$ mixing of
Lenz and Nierste \cite{Ulrich} and references therein.

$B_{s}-\bar{B}_{s}$ oscillations can be studied by
Schr\" odinger equation

\begin{eqnarray}
\imath \frac{d}{dt}\left( \begin{array}{c} |B_{s}(t)\rangle \\
|\bar{B}_{s}(t)\rangle \end{array} \right) =
 \left( M^{s} - \frac{\imath}{2} \Gamma^{s} \right)
\left( \begin{array}{c} |B_{s}(t)\rangle \\
|\bar{B}_{s}(t)\rangle \end{array} \right)
\end{eqnarray}

where the mass matrix is $M^{s}$ and the decay matrix $\Gamma^{s}$. 
The physical eigenstates $|B_{H}\rangle$ and $|B_{L}\rangle$
are achieved by
diagonalization of the matrix $M^{s}-\frac{\imath}{2}\Gamma^{s}$.
We can write,
to a very good approximation, for the mass and
width differences \cite{Ulrich}

\begin{eqnarray}
\Delta M_{s} = M^{s}_{H} - M^{s}_{L} = 2 |M^{s}_{12}|,\ 
\Delta \Gamma_{s} = \Gamma^{s}_{L} - \Gamma^{s}_{H} = 
 2 |\Gamma^{s}_{12}| \cos \phi_{s}, 
\end{eqnarray}
\begin{eqnarray*}
CP\ phase\ \phi_{s} = arg(-M^{s}_{12}/\Gamma^{s}_{12}). 
\end{eqnarray*}

It is precisely the off-diagonal mass matrix element $M^{s}_{12}$ 
that is
potentially the most sensitive quantity to new physics because
it is defined by the quantum loop, i.e. a box diagram. The off-diagonal
part of the decay matrix $\Gamma^{s}_{12}$ is dominated by 
the tree level processes.

The SM prediction for $M^{s}_{12}$ is well known \cite{Ulrich}

\begin{eqnarray}
M_{12}^{s} = \frac{G_{F}^{2}M_{B_{s}}}{12\pi^{2}}
M_{W}^{2}(V_{tb}V_{ts}^{*})^{2}\hat{\eta}_{B}S_{0}(x_{t})
f_{B_{s}}^{2} B, \hspace{50mm} \\
G_{F}=Fermi\ constant,\ V_{ij}=CKM\ matrix\ elements, \nonumber \\  
M_{B_{s}}=mass\ of\ B_{s}\ meson,\ M_{W}=W\ boson\ mass,\nonumber \\
\ \hat{\eta}_{B}=QCD\ correction\ factor,\ 
S_{0}(x_{t})=Inami-Lim\ function,\nonumber \\ x_{t}=m_{t}^{2}/M_{W}^{2},\
m_{t}=t-quark\ mass, \nonumber
\end{eqnarray}
\begin{eqnarray*}
\langle B_{s}| Q | \bar{B}_{s} \rangle = \frac{8}{3}
M_{B_{s}}^{2} f_{B_{s}}^{2} B, \hspace{50mm} \\
Q = \bar{s}_{\alpha}\gamma_{\mu}(1-\gamma_{5})b_{\alpha}
\bar{s}_{\beta}\gamma^{\mu}(1-\gamma_{5})b_{\beta},
\ \alpha,\beta=1,2,3=colour\ indices.
\end{eqnarray*}

Lenz and Nierste \cite{Ulrich} improved essentially the theoretical 
prediction for $\Gamma^{s}_{12}$ by introducing more a natural operator
basis and resumming charm quark contributions.

Now, we inspect where we can expect the
largest deviation of the BY theory from the SM. Apart from
the nonperturbative estimates of the matrix elements with hadrons,
one encounters also QCD corrections, but at the scale of
$\mu \simeq m_{b}$, the difference between the SM and BY strong
couplings is negligible \cite{Palle6}. The largest deviation
is expected at short distance contributions when calculating
box diagrams with the heaviest internal particles.

The SM formulas of the box diagrams (two virtual W bosons and two 
virtual u-, c- or t-quarks) responsible for
$K-\bar{K}$, $B_{d}-\bar{B_{d}}$ or $B_{s}-\bar{B_{s}}$
mixings are provided by Inami and Lim \cite{Inami}.
A pedagogical derivation of the box diagrams in $R_{\xi}$ gauges
for vanishing external masses and momenta can be found
in \cite{Silva}. The final gauge invariant result 
looks like the following:

\begin{eqnarray}
Box\ diagram\ (Fig.17.2;\ Ref.[19])=\frac{g^{4}}{4}(\bar{s_{1}}
\Gamma^{\mu}d_{1})(\bar{s_{2}}\Gamma_{\mu}d_{2}) \nonumber \\
\times 
\sum_{\alpha=u,c,t}\sum_{\beta=u,c,t} \lambda_{\alpha}\lambda_{\beta}
\int\frac{d^{4}k}{(2\pi)^{4}} 
\frac{1}{D_{\alpha}D_{\beta}D_{W}^{2}}
\left(\frac{k^{2}m_{\alpha}^{2}m_{\beta}^{2}}{4 M_{W}^{4}}
+ k^{2} -\frac{2 m_{\alpha}^{2}m_{\beta}^{2}}{M_{W}^{2}} \right),
\end{eqnarray}
\begin{eqnarray*}
D_{\alpha} &\equiv& k^{2} - m_{\alpha}^{2},\ 
D_{W} \equiv k^{2} - M_{W}^{2},\ \Gamma_{\mu} \equiv \gamma_{\mu}
\frac{1}{2}(1-\gamma_{5}),  \hspace{45mm} \\
\lambda_{\alpha}&\equiv& V^{*}_{\alpha s}V_{\alpha d}\ for\ 
K^{0}-\bar{K^{0}}\ system;
\lambda_{\alpha}\equiv V^{*}_{\alpha b}V_{\alpha d}\ for\ 
B^{0}_{d}-\bar{B^{0}_{d}}\ system; \\
\lambda_{\alpha}&\equiv& V^{*}_{\alpha b}V_{\alpha s}\ for\ 
B^{0}_{s}-\bar{B^{0}_{s}}\ system.
\end{eqnarray*}

Resolving the UV convergent integral in the Feynman fashion,
one finds \cite{Silva}

\begin{eqnarray}
Box\ diagram = \frac{\imath g^{4}}{64\pi^{2}M_{W}^{2}}
(\bar{s_{1}}\Gamma^{\mu}d_{1})(\bar{s_{2}}\Gamma_{\mu}d_{2})
\sum_{\alpha=u,c,t}\sum_{\beta=u,c,t} \lambda_{\alpha}\lambda_{\beta}
F(x_{\alpha},x_{\beta}), \nonumber \\
\hspace{10 mm}
\end{eqnarray}
\begin{eqnarray}
F(x_{\alpha},x_{\beta})\equiv \frac{1}{(1-x_{\alpha})(1-x_{\beta})}
\left(\frac{7x_{\alpha}x_{\beta}}{4}-1\right) 
+ \frac{x_{\alpha}^{2}\ln x_{\alpha}}{(x_{\beta}-x_{\alpha})
(1-x_{\alpha})^{2}} \nonumber \\ \times \left(1-2 x_{\beta} + 
\frac{x_{\alpha}x_{\beta}}{4}\right) 
   +\frac{x_{\beta}^{2}\ln x_{\beta}}{(x_{\alpha}-x_{\beta})
(1-x_{\beta})^{2}}\left(1-2 x_{\alpha} + 
\frac{x_{\alpha}x_{\beta}}{4}\right), \nonumber \\ 
x_{\alpha} \equiv \frac{m_{\alpha}^{2}}{M_{W}^{2}}.\nonumber
\end{eqnarray}

This form can be simplified by the unitarity of the CKM matrix
$\lambda_{u}+\lambda_{c}+\lambda_{t}=0$ 
and with the approximation $m_{u}=0$, we get \cite{Silva}:

\begin{eqnarray}
Box\ diagram &=&\frac{-\imath G_{F}^{2}M_{W}^{2}}{2\pi^{2}}
(\bar{s_{1}}\Gamma^{\mu}d_{1})(\bar{s_{2}}\Gamma_{\mu}d_{2}) 
{\cal F}_{0}, \\  
{\cal F}_{0}&=&\lambda_{c}^{2}S_{0}(x_{c})+\lambda_{t}^{2}S_{0}(x_{t})
+2 \lambda_{c}\lambda_{t}S_{0}(x_{c},x_{t}), \nonumber \\
S_{0}(x_{c},x_{t})&=&-F(x_{c},x_{t})-F(0,0)+F(0,x_{c})+
F(0,x_{t}), \nonumber \\
S_{0}(x_{c})&=&\lim_{x_{t}\rightarrow x_{c}} S_{0}(x_{c},x_{t}),
\nonumber \\
\ S_{0}(x)&=&\frac{x}{(1-x)^{2}}[1-\frac{11x}{4}+\frac{x^{2}}{4}
-\frac{3x^{2}\ln x}{2(1-x)}]. \nonumber
\end{eqnarray}

The dominant t-quark short distance contribution in the effective quark
theory for the mass difference $M^{s}_{12}$ enters in the form of 
the Inami-Lim function $S_{0}(x_{t})$.

The Nambu-Goldstone scalars in the BY theory are decoupled from
quarks at the tree level \cite{Palle1} and they carry lepton
number.
Thus, one has to perform calculations of the box diagrams
in the unitary gauge. The form of the subintegral
function in Eq.(4) is the same as in the SM because of the gauge
invariance($\xi \rightarrow 0$ in the unitary gauge:
$\Delta_{\mu\nu}=-\imath (g_{\mu\nu}-\frac{p_{\mu}p_{\nu}}{M^{2}})
\frac{1}{p^{2}-M^{2}}-\imath \frac{p_{\mu}p_{\nu}}{M^{2}}
\frac{1}{p^{2}-M^{2}/\xi},\ \Delta=\frac{\imath}{p^{2}-M^{2}/\xi}$).
If it is easier to calculate an observable in $R_{\xi}$ rather than
in the unitary gauge, it is allowed to do so even in the BY theory
because of the gauge invariance (i.e. independence on the $\xi$
parameter) of the observable. The perturbation theory with
coupling as a perturbation parameter in the
strong coupling system of the heavy Majorana neutrinos and the
Nambu-Goldstone scalars is useless and Dyson-Schwinger equations
within the nonsingular BY theory are a framework for the analysis. 

Let us divide the UV convergent integral of the box diagram Eq. (4)
into two UV convergent integrals introducing the UV cut-off 
after Wick's rotation.
These integrals can be evaluated numerically or analitically by
elementary functions (albeit with lengthy expressions) to recheck
their sum evaluated previously in Eq. (5):  

\begin{eqnarray}
F(x_{\alpha},x_{\beta})&\equiv& F^{\infty}(x_{\alpha},x_{\beta})
= F^{\Lambda}(x_{\alpha},x_{\beta}) -
\Delta F^{\Lambda}(x_{\alpha},x_{\beta}), \hspace{30mm} \\
F^{\Lambda}(x_{\alpha},x_{\beta}) &=& -M_{W}^{2}\int^{\Lambda^{2}}_{0}
dz z(z+m_{\alpha}^{2})^{-1}(z+m_{\beta}^{2})^{-1}
(z+M_{W}^{2})^{-2} \nonumber \\
&\times& \left(z(1+\frac{m_{\alpha}^{2}m_{\beta}^{2}}
{4 M_{W}^{4}}) + \frac{2m_{\alpha}^{2}m_{\beta}^{2}}{M_{W}^{2}}\right),
\nonumber \\
\Delta F^{\Lambda}(x_{\alpha},x_{\beta}) &=& +M_{W}^{2}
\int^{1/\Lambda^{2}}_{0} dw (1+m_{\alpha}^{2}w)^{-1}
 (1+m_{\beta}^{2}w)^{-1}(1+m_{W}^{2}w)^{-2} \nonumber \\
&\times&\left(1 + \frac{m_{\alpha}^{2}m_{\beta}^{2}}{4 M_{W}^{4}}
+ 2\frac{m_{\alpha}^{2}m_{\beta}^{2}}{M_{W}^{2}}w\right).
\nonumber
\end{eqnarray}

Thus, the BY theory with the universal UV cut-off defined by
the weak boson mass $\Lambda=326 GeV$ predicts for 
$M^{s}_{12}$ ($\hat{\eta}_{B}$ QCD correction function in the BY theory 
remains almost unchanged with respect to the SM:$\mu \simeq m_{b}
<< \Lambda $ )
owing to the electroweak box diagram (see Eq.(3) and Eqs.(5-7)): 

\begin{eqnarray}
M^{s}_{12}(BY) &=& \frac{S_{0}^{\Lambda}(x_{t})}{S_{0}^{\infty}(x_{t})}
M^{s}_{12}(SM), \hspace{50mm} \\
S_{0}^{\Lambda}(x_{c},x_{t})&\equiv&-F^{\Lambda}(x_{c},x_{t})
-F^{\Lambda}(0,0)+F^{\Lambda}(0,x_{c})+
F^{\Lambda}(0,x_{t}), \nonumber \\
S_{0}^{\Lambda}(x_{t})&=& \lim_{x_{c}\rightarrow x_{t}}
S_{0}^{\Lambda}(x_{c},x_{t}). \nonumber
\end{eqnarray}

The final relation for the CP asymmetry in the flavour-specific
$B_{s} \rightarrow f$ decays has the form \cite{Ulrich}:

\begin{eqnarray}
a_{fs}^{s} = \frac{|\Gamma^{s}_{12}|}{|M^{s}_{12}|}\sin \phi_{s}.
\end{eqnarray}

In the last chapter, we present results and concluding remarks. 

\section{Results and conclusions}

Lenz and Nieste \cite{Ulrich} reported in 2007 2$\sigma$ deviation
for the CP violating phase $\phi_{s}$ despite of the fact that
the mass difference $M^{s}_{12}$ (Eq.(3)) contains poorly
known $V_{CKM}$ matrix elements and $f^{2}_{B_{s}}B$ form factor.
Hadron models still generate the largest uncertainty in studying
semileptonic and hadronic processes where matrix elements are
extracted: $\mid V_{td}\mid=(8.4\pm 0.6)
\cdot 10^{-3},\ \mid V_{ts}\mid=(38.7\pm 2.1)\cdot 10^{-3},
\mid V_{tb}\mid=0.88\pm 0.07$ \cite{PDG}.
For the detailed error budget analysis the reader can consult
Ref.\cite{Ulrich}.

We can easily estimate the deviation of the BY theory prediction
from the SM one. From the preceding chapter, one concludes that
only the short distance electroweak part of $M^{s}_{12}$, 
hidden in the modified Inami-Lim function, can enhance the
magnitude of the semileptonic CP asymmetry parameter \cite{Ulrich}: 

\begin{eqnarray}
a_{sl}(SM) &\simeq & 0.582 a_{sl}^{d} + 0.418 a_{sl}^{s},\ \nonumber \\
M_{W} &=& 80.4\ GeV,\ m_{c}=1.3\ GeV,\ m_{t}=172\ GeV \nonumber \\
&\Rightarrow & 
\kappa \equiv a_{sl}(BY)/a_{sl}(SM) =S_{0}^{\infty}(x_{t})
/S_{0}^{\Lambda}(x_{t})=1.13 .
\end{eqnarray}

\rotatebox{-90}{
\epsfig{figure=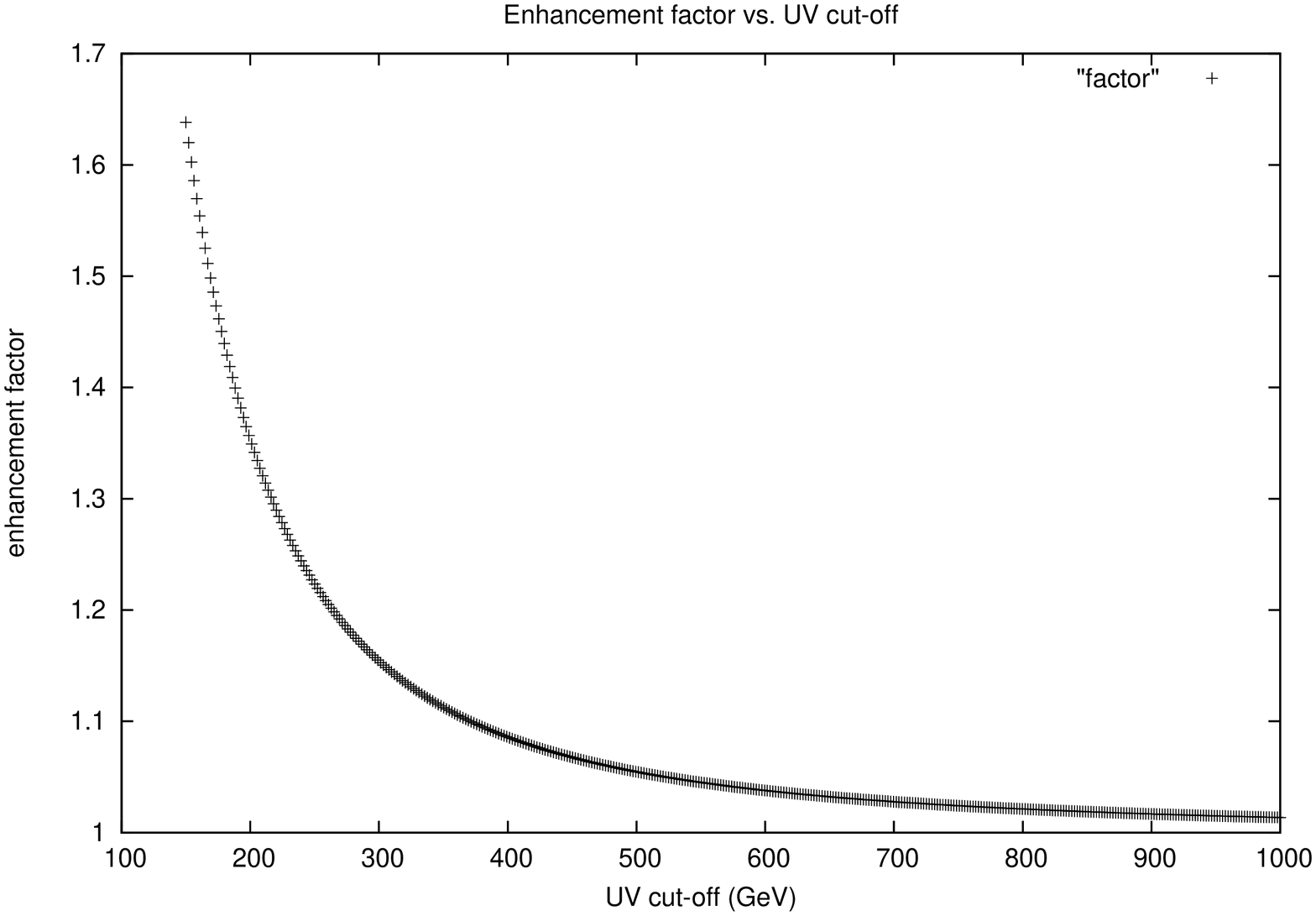, height=130 mm, width=90 mm}}

\vspace{25mm}

{\bf Fig. 1: Enhancement factor $\kappa$ as a function of
the UV cut-off (GeV).}
\newline

In Fig. 1, the reader can visualize the enhancement factor as 
a function of the UV cut-off, however one should bear in mind
that $\Lambda$ is not a free parameter in the BY theory.

The larger magnitude of the asymmetry
parameter for the BY theory by $13\%$ is still significantly smaller than
the median value of the experiment \cite{D0}.
In numerical evaluations of the CP asymmetry parameter of the $B_d$
system, more intermediate experimental values are incorporated than 
for $B_s$ system. However, the BY theory predicts the same enhancement
for both systems.

It is important to notice that the mixing of the neutral
kaons is also affected by the virtual heavy t-quark. We can 
estimate the deviation from the SM (see ch. 17 of Ref. \cite{Silva}
for the theory and Ref. \cite{Amsler} for Wolfenstein parameters):

\begin{eqnarray}
M_{12}&\propto&\Re (\lambda_{c}^{*}\lambda_{u})
[\eta_{1}S_{0}(x_{c})-\eta_{3}S_{0}(x_{c},x_{t})] \nonumber \\
&+&\Re (\lambda_{t}^{*}\lambda_{u})
[\eta_{3}S_{0}(x_{c},x_{t})-\eta_{2}S_{0}(x_{t})],  \\
&&QCD\ correction\ factors:\ \eta_{1}=1.38,\ \eta_{2}=0.57,\ 
\eta_{3}=0.47, \nonumber 
\end{eqnarray}
\begin{eqnarray}
V_{CKM}=\left( \begin{array}{ccc}
1-\lambda^{2}/2 & \lambda & A\lambda^{3}(\rho-\imath \eta ) \\
-\lambda & 1-\lambda^{2}/2 & A\lambda^{2} \\
A\lambda^{3}(1-\rho-\imath \eta) & -A\lambda^{2} & 1
\end{array} \right) + {\cal O}(\lambda^{4}), \hspace{66mm} , \nonumber
\end{eqnarray}
\begin{eqnarray}
Wolfenstein\ parameters\ for\ V_{CKM}:\ 
\lambda=0.2257,\ A=0.814, \nonumber \\ \rho=0.135,\ \eta=0.349, \nonumber \\
S_{0}^{\infty}(x_{t})/S_{0}^{\Lambda}(x_{t})=1.13,\
S_{0}^{\infty}(x_{c},x_{t})/S_{0}^{\Lambda}(x_{c},x_{t})=1.014,\nonumber \\  
S_{0}^{\infty}(x_{c})/S_{0}^{\Lambda}(x_{c})=1.000 \nonumber \\
\Rightarrow  |M_{12}(SM)/M_{12}(BY)| = 1.055 .\hspace{40mm}
\end{eqnarray}

The most recent analysis of Lenz \cite{Lenz2} based on the new LHCb measurements \cite{LHCb}
rules out large SM deviation suggested by Tevatron data. Moreover, it strengthens the 
validity of the heavy quark expansion corrections (HQE) to the amplitudes \cite{Lenz2}:

\begin{eqnarray}
(\frac{\Delta \Gamma_{s}}{\Delta M_{s}})^{Exp}/
(\frac{\Delta \Gamma_{s}}{\Delta M_{s}})^{SM} = 1.12 \pm 0.27
\end{eqnarray}

\hspace{60 mm} and 

\begin{eqnarray}
\frac{\Delta \Gamma_{s}^{Exp}}
{\Delta \Gamma_{s}^{SM}} = 1.15\pm 0.32 .
\end{eqnarray}

 The theory relies on the HQE, the QCD corrections and lattice gauge theory 
evaluations of the hadron matrix elements. The total discrepancy between the theory
and the experiment at the level of $30\%$ \cite{Lenz2}
does not exclude small short distance corrections of the BY theory.
Since the BY theory is UV finite, the extensive use of the Dyson-Schwinger and
Bethe-Salpeter amplitudes and equations can reassure lattice and HQE calculations.
The dependence of QCD amplitudes in the strong coupling regime on the UV cutoff
is only logarithmic or log-log when solving the bootstrap equations. The essential dependence of the solutions 
appears in the parametrization of the infrared sector \cite{Jain} that
does not differ from the SM one.

To conclude, the noncontractible spacetime as a symmetry breaking
mechanism in the BY theory of Ref. \cite{Palle1} changes the SM
processes essentially via quantum loops if internal masses or 
external momenta and masses are large \cite{Palle6}. We show
that the enhanced baryon CP violation in the BY theory cannot explain
Tevatron results \cite{D0}, but its prediction can help to resolve
some present and possible future discrepancy of the
unitarity relations of
the mixing matrix. It seems that new measurements at the LHCb
\cite{LHCb} of $B_{s} \rightarrow J/\Psi f_{0}(980)\ 
and\ \Phi$ do not support a large deviation 
from the SM prediction.
It could be more favourable to search for the electroweak deviations of the 
SM in the $B_{s} \rightarrow \mu \mu$ decay \cite{Palle8}
where the hadron physics uncertainty is smaller.

\end{document}